\documentclass[%
letterpaper,
final,
tightenlines,
raggedbottom,
showpacs,
twoside,
twocolumn,
pdfstartview=FitH,
10pt,
amsmath,amssymb,amsfonts,
aps,
pra,
floatfix,
]{revtex4-1}

\input{glyphtounicode}
\newcommand{\footnoteremember}[2]{
\footnote{#2}\newcounter{#1}\setcounter{#1}{\value{footnote}}}
\newcommand{\footnoterecall}[1]{
\footnotemark[\value{#1}]}

\usepackage[hyphenbreaks]{breakurl}

\usepackage[sort&compress]{natbib}
\usepackage[
	bookmarks,
	breaklinks,
	colorlinks,
	allcolors=blue,
	urlcolor=blue,
	linktocpage,
	hyperfootnotes,
	bookmarkstype=toc,
	bookmarksopen,
	bookmarksopenlevel=1
]{hyperref}
\hypersetup{pdfpagemode=UseNone}

\usepackage{textcomp}
\usepackage{graphicx}
\usepackage{dcolumn}
\usepackage{bm}

\usepackage[all]{hypcap}

\usepackage{xcolor}
\definecolor{grey}{rgb}{0.5,0.5,0.5}
\definecolor{lightgrey}{rgb}{0.9,0.9,0.9}

\makeatletter
\renewcommand*{\p@section}{}
\renewcommand*{\p@subsection}{}
\renewcommand*{\p@subsubsection}{}
\makeatother

\begin{document}

\title{Ensemble master equation for a trapped-atom clock with one- and two-body losses}


\author{Valentin Ivannikov}
\altaffiliation[Present address: ]{Physikalisches Institut, Universit\"at Heidelberg, Germany}
\email{ivannikov@physi.uni-heidelberg.de}
\affiliation{Centre for Atom Optics and Ultrafast Spectroscopy, Swinburne University of Technology, Melbourne, Australia}


\date{14 February 2014}

\begin{abstract}
An ensemble density matrix model that includes one- and two-body losses is derived for a trapped-atom clock. A trapped-atom clock is mainly affected by one- and two-body losses, generally giving nonexponential decays of populations; nevertheless, three-body recombination is also quantitatively analyzed to demonstrate the boundaries of its practical relevance. The importance of one-body losses is highlighted without which population trapping behavior would be observed. The model is written with decay constants expressed through experimental parameters. It can complement, e.g., the ISRE (identical spin rotation effect) model to improve its predictions: ISRE dramatically increases the ensemble coherence time, hence it enables one to observe the influence of two-body losses on the interferometry contrast envelope. The presented model is useful for Ramsey interferometry and is ready for immediate experimental verification in existing systems.
\end{abstract}

\pacs{67.85.--d, 06.20.fb, 06.30.Ft, 37.10.Gh}

\maketitle

\section{Introduction}

An atom clock has become the ultimate accurate experimental tool since the invention of the method of separated oscillatory fields by Ramsey in 1949 \cite{Ramsey1949}. During the course of technological advancement the miniaturization of the atom clock has become a vibrant topic in navigation, space positioning systems, communications, sensing applications, and quantum memories \cite{Swallows2010,Hinkley2013,KleineBuning2011,Hammerer2010} for which trapped-atom clocks are being developed \cite{Treutlein2004,Riedel2010,Ramirez-Martinez2011,Rosenbusch2009,KleineBuning2011}. However, trapping ensues high atomic densities whose many-body interactions via collisions of atoms in the clock states become important \cite{Martin2013}. In the usual experimental settings the atomic density of an ensemble reaches values where two-body interactions can lead to apparent losses. Then ensemble evolution equations acquire additional complexity via the inclusion of one- and two-body losses that make exact analytical solution prohibitive in treatment. For this reason one- or two-body loss only is typically included \cite{Deutsch2010a,Maineult2012,KleineBuning2011,Tojo2009,Mertes2007,Egorov2012,Martin2013}. However, achievable vacuum is imperfect and differs noticeably between systems \cite{Knoop2012}, and the actual pressures during data acquisition range from $10^{-11}$ to $10^{-9}$~Torr, influencing the one-body decay rates. Albeit, two-body losses alone lead to population trapping [\citep[p.~102]{Ivannikov2013thesis} and Eqs.~(\ref{eq:finalwithlosses})]. This implies that one-body processes should be included in the model to avoid consequential errors in measurement interpretation.

In this work an ensemble master equation with one- and two-body inelastic collisional losses and a phenomenological dephasing is derived by employing the open systems approach and including interactions of the system with population and phase damping reservoirs. Along the way a set of relevant population decay laws can be extracted for a comprehensive atom number relaxation analysis as special cases of the ensemble master equation.

We first introduce the measurables $N_1$, $N_2$, and $P_z$ usually sought in an experiment in terms of the ensemble density matrix elements and define the transition from atomic densities to numbers of atoms. Then the interaction with the bath is described in the Markovian framework to account for the loss of coherence associated with the particle loss. Further, we introduce the phenomenological dephasing $\gamma_d$ (and its counterpart $\Gamma_d$ in the ensemble master equation) to account for the phase differences that are not related to the population loss. Finally, the ensemble master equation [Eqs.~(\ref{eq:finalwithlosses})] is written with the decay constants defined purely in terms of experimental parameters.

\section{Master equation with many-body losses}

The measurables are usually the normalized populations $N_1/N$, $N_2/N$, where $N = N_1 + N_2$, or the normalized population difference $P_z$. In terms of atom numbers $N_{j}$ and the ensemble density matrix elements $\rho_{jj}$ with state index $j$, $P_z$ can be expressed as
\begin{equation}
P_z = \frac{N_1 - N_2}{N} = \frac{\rho_{11} - \rho_{22}}{\rho_{11} + \rho_{22}}.
\label{eq:measurables}
\end{equation}
To compare the model with the experimental observations of $N_1$ and $N_2$, the thermal cloud densities are integrated over space: $N_j(t) = \int \! n_j(t,\text{\textbf{\textit{r}}}) \, d^3\text{\textbf{\textit{r}}}$.

One-body losses, that can be measured, for example, in a magnetic trap, are a sign of imperfect vacuum: hot background atoms collide with the trapped ensemble transferring enough momentum to the cold particles for them to escape from the trap. The master equation is parametrized in this case by the one-body decay rates $\Gamma_{1}$ and $\Gamma_{2}$ for the respective states. Nonetheless, there is experimental evidence of a strong influence of two-body inelastic collisions on the decay of the populations \cite{Widera2005}. To incorporate many-body collisions into the model, the interaction of a field of bosons with damping reservoirs needs to be considered. For this, a suitable Markovian model setup is adopted from \cite{Opanchuk2012a} and methods from \cite{Tojo2009,Carmichael1993book} are used in the following derivations. It is also assumed that the particles that experience inelastic collisions immediately leave the trap without interaction with the rest of the ensemble.

Most generally, the system-bath interaction is described by the Markovian master equation:
\begin{equation}
\frac{\partial\hat\rho}{\partial t} = \frac{1}{i\hbar}\left[\hat{\mathbf{H}},\hat\rho\right] + \sum_{\text{\textbf{\textit{b}}}\in\mathbb{B}}{\mathcal{G}_{\text{\textbf{\textit{b}}}}\int \! \mathcal{\hat{L}}_{\text{\textbf{\textit{b}}}}\left[\hat\rho\right] \, d^3\text{\textbf{\textit{r}}}},
\label{eq:markov}
\end{equation}
where $\hat{\mathbf{H}}$ is the system Hamiltonian, the commutator describes the coherent evolution of the system, the interaction part is further referred to as $\left(\left.\partial\hat\rho\right/\partial t\right)_{\text{loss}}$, and $\mathbb{B} = \left\{\left(b_1,b_2, ...,b_S\right)\right\}$ is a set of tuples each with a number of elements corresponding to a number of states $S$ interacting via simultaneous collisions where each element contains a number of colliding particles in a given state. Index $\text{\textbf{\textit{b}}}$ chooses one tuple $\left(b_1,b_2, ...,b_j, ...,b_S\right)$ that describes all particles in all states interacting with the tuple loss channel characterized by the constant $\mathcal{G}_{\text{\textbf{\textit{b}}}}$. The set element $b_j$ is the number of colliding particles in state $j$. The cardinality of the set $\left\vert \mathbb{B} \right\vert$ equals the number of loss channels. Interaction with reservoirs via loss channel $\text{\textbf{\textit{b}}}$ is found by applying the Lindblad superoperator to the density operator:
\begin{equation}
\mathcal{\hat{L}}_{\text{\textbf{\textit{b}}}}\left[\hat\rho\right] = 2\hat{O}_{\text{\textbf{\textit{b}}}}\hat\rho\hat{O}_{\text{\textbf{\textit{b}}}}^\dagger - \hat{O}_{\text{\textbf{\textit{b}}}}^\dagger\hat{O}_{\text{\textbf{\textit{b}}}}\hat\rho - \hat\rho\hat{O}_{\text{\textbf{\textit{b}}}}^\dagger\hat{O}_{\text{\textbf{\textit{b}}}}.
\label{eq:superop}
\end{equation}
The operators $\hat{O}_{\text{\textbf{\textit{b}}}}$ define the collision of all of the particles in all of the states defined by the corresponding tuple in terms of the field operators:
\begin{equation}
\hat{O}_{\text{\textbf{\textit{b}}}} = \prod^{S}_{j=1} \hat\Psi^{b_j}_{j}(\text{\textbf{\textit{r}}}).
\end{equation}
The field operators are expanded in terms of mode functions $\psi_{jk}(\text{\textbf{\textit{r}}})$ and single-mode operators as \cite{Norrie2006a}
\begin{equation}
\hat\Psi_j(\text{\textbf{\textit{r}}}) = \sum_{k} \hat a_{jk} \psi_{jk}(\text{\textbf{\textit{r}}}),
\label{eq:fieldop}
\end{equation}
where $\hat a_{jk}$ is an operator that destroys a particle in state $j$, momentum mode $k$, and obeys bosonic commutation relations. The arguments $\text{\textbf{\textit{r}}}$ will be omitted hereafter for compactness: $\hat\Psi_j \equiv \hat\Psi_j(\text{\textbf{\textit{r}}})$. The field operators are defined in the standard way for creation $\hat\Psi^\dagger_j(\text{\textbf{\textit{r}}})$ and annihilation $\hat\Psi_j(\text{\textbf{\textit{r}}})$ in state $j \in \{1,2, ..., S\}$ obeying the commutation relation for identical bosons:
\begin{equation}
[\hat\Psi_i(\text{\textbf{\textit{r}}}), \hat\Psi^\dagger_j(\text{\textbf{\textit{r}}}')] = \delta_{ij}\delta(\text{\textbf{\textit{r}}}-\text{\textbf{\textit{r}}}').
\label{eq:bosoncomm}
\end{equation}

The unitary and the loss parts of Eq.~(\ref{eq:markov}) can be treated independently. To compute a measurable corresponding to the interaction part of Eq.~(\ref{eq:markov}), averages of the operator elements should be found by evaluating the trace:
\begin{equation}
\frac{\partial}{\partial t}\langle \hat\Psi_i^\dagger\hat\Psi_j \rangle = \operatorname{Tr}\left[\left(\frac{\partial\hat\rho}{\partial t}\right)_{\text{loss}}\hat\Psi_i^\dagger\hat\Psi_j\right].
\label{eq:trace}
\end{equation}
By calculating the expectation values from Eq.~(\ref{eq:trace}) we arrive at the following system of equations:
\begin{subequations}
\begin{align}
	\frac{\partial}{\partial t}\langle \hat\Psi_1^\dagger\hat\Psi_1 \rangle &= -2\mathcal{G}_{(1,0)}\langle \hat\Psi_1^\dagger\hat\Psi_1 \rangle -2\mathcal{G}_{(1,1)}\langle \hat\Psi_1^\dagger\hat\Psi_1 \hat\Psi_2^\dagger\hat\Psi_2 \rangle \nonumber\\
	&\quad - 4\mathcal{G}_{(2,0)}\langle \hat\Psi_1^{\dagger 2}\hat\Psi_1^2 \rangle,\\
	\frac{\partial}{\partial t}\langle \hat\Psi_2^\dagger\hat\Psi_2 \rangle &= -2\mathcal{G}_{(0,1)}\langle \hat\Psi_2^\dagger\hat\Psi_2 \rangle -2\mathcal{G}_{(1,1)}\langle \hat\Psi_1^\dagger\hat\Psi_1 \hat\Psi_2^\dagger\hat\Psi_2 \rangle \nonumber\\
	&\quad - 4\mathcal{G}_{(0,2)}\langle \hat\Psi_2^{\dagger 2}\hat\Psi_2^2 \rangle,\\
	\frac{\partial}{\partial t}\langle \hat\Psi_1^\dagger\hat\Psi_2 \rangle &= -\mathcal{K}_1\langle \hat\Psi_1^\dagger\hat\Psi_2 \rangle - \mathcal{K}_2\langle \hat\Psi_1^\dagger\hat\Psi_2 \hat\Psi_1^\dagger\hat\Psi_1 \rangle \nonumber\\
	&\quad - \mathcal{K}_3\langle \hat\Psi_2^\dagger\hat\Psi_2 \hat\Psi_1^\dagger\hat\Psi_2 \rangle,\\
	\frac{\partial}{\partial t}\langle \hat\Psi_2^\dagger\hat\Psi_1 \rangle &= -\mathcal{K}_1\langle \hat\Psi_2^\dagger\hat\Psi_1 \rangle - \mathcal{K}_2\langle \hat\Psi_1^\dagger\hat\Psi_1 \hat\Psi_2^\dagger\hat\Psi_1 \rangle \nonumber\\
	&\quad- \mathcal{K}_3\langle \hat\Psi_2^\dagger\hat\Psi_1 \hat\Psi_2^\dagger\hat\Psi_2 \rangle,
\end{align}%
\label{eq:mbloss}%
\end{subequations}%
where the reservoir constants are absorbed in the following scalars for compactness: $\mathcal{K}_1 = \mathcal{G}_{(1,0)}+\mathcal{G}_{(0,1)}$, $\mathcal{K}_2 = 2\mathcal{G}_{(2,0)}+\mathcal{G}_{(1,1)}$, and $\mathcal{K}_3 = 2\mathcal{G}_{(0,2)}+\mathcal{G}_{(1,1)}$. The field operators are ordered suitably for the physical interpretation: The average $\langle \hat\Psi_j^\dagger\hat\Psi_j \rangle$ corresponds to the density of particles in state $j$ at the position defined by the coordinate vector $\text{\textbf{\textit{r}}}$. The averages $\langle \hat\Psi_i^\dagger\hat\Psi_j \rangle$ with $i\neq j$ carry phase difference information between states $i$ and $j$. Now that we see the identification of the involved processes, the structure of Eq.~(\ref{eq:mbloss}c), for example, shows there are three independent terms $\langle \hat\Psi_1^\dagger\hat\Psi_2 \rangle$, $\langle \hat\Psi_1^\dagger\hat\Psi_2 \hat\Psi_1^\dagger\hat\Psi_1 \rangle$, and $\langle \hat\Psi_2^\dagger\hat\Psi_2 \hat\Psi_1^\dagger\hat\Psi_2 \rangle$ responsible for damping to the associated population reservoirs. Each of the prefactors to the operator averages on the right-hand side has a clear physical meaning as a decay rate. We can establish an identification of $\mathcal{G}_{\text{\textbf{\textit{b}}}}$ to the corresponding decay rates by letting $2\mathcal{G}_{(1,0)} \!\rightarrow\! \gamma_1$, $2\mathcal{G}_{(0,1)} \!\rightarrow\! \gamma_2$, $2\mathcal{G}_{(1,1)} \!\rightarrow\! \gamma_{12}$, $4\mathcal{G}_{(2,0)} \!\rightarrow\! \gamma_{11}$, and $4\mathcal{G}_{(0,2)} \!\rightarrow\! \gamma_{22}$, where on the right-hand side we have decay rates. To distinguish the decay rates measured with thermal atoms and Bose-Einstein condensate (BEC), later in the text they are superscripted with ``th'' and ``BEC,'' correspondingly, and the identity $\gamma_{\mathbf{\ell}} \!\equiv\! \gamma_{\mathbf{\ell}}^{\text{th}}$ holds.

In the mean-field approximation the loss part of Eq.~(\ref{eq:markov}) $\left(\left.\partial\hat\rho\right/\partial t\right)_{\text{loss}}$ with $\kappa_1 \!\!=\!\! \frac{\gamma_{1}+\gamma_{2}}{2}$, $\kappa_2 \!=\!\! \frac{\gamma_{11}+\gamma_{12}}{2}$, and $\kappa_3 \!=\! \frac{\gamma_{22}+\gamma_{12}}{2}$ simplifies to the equations for the atomic densities:
\begin{equation}
\begin{aligned}
	\frac{\partial n_{1}}{\partial t} &= -\gamma_{1} n_{1} -\gamma_{12} n_{1} n_{2} - \gamma_{11} n_{1}^2,\\
	\frac{\partial n_{2}}{\partial t} &= -\gamma_{2} n_{2} -\gamma_{12} n_{1} n_{2} - \gamma_{22} n_{2}^2,\\
	\frac{\partial n_{12}}{\partial t} &= -\kappa_1 n_{12} - \kappa_2 n_{1} n_{12} - \kappa_3 n_{2} n_{12},\\
	\frac{\partial n_{21}}{\partial t} &= -\kappa_1 n_{21} - \kappa_2 n_{1} n_{21} - \kappa_3 n_{2} n_{21}.
\end{aligned}
\label{eq:mbatdensity}%
\end{equation}

\section{Phenomenological dephasing}
\label{sec:phendeph}

Collisional dephasing caused by population loss appears in the off-diagonal density operator elements naturally while accounting for inelastic collisions between particles. However, there are other mechanisms of dephasing not related to population loss, e.g., elastic collisions or inhomogeneity of the trapping potential. Such \textit{pure dephasing} can be introduced phenomenologically by adding other reservoirs for phase damping of state $1$ $R_{\text{ph}1}$, state $2$ $R_{\text{ph}2}$, etc., to the total Hilbert space by combining the subspaces of the system $S$, particle damping reservoir $R$, phase damping reservoirs as well as the coupling degrees of freedom that describe interactions between them as $S \oplus R \oplus SR \oplus R_{\text{ph}1} \oplus R_{\text{ph}2} \oplus SR_{\text{ph}1} \oplus SR_{\text{ph}2}$. The Lindblad superoperator should then only couple the states to $R_{\text{ph}1}$ and $R_{\text{ph}2}$, since states are \textit{phase sources}. A suitable operator reads
\begin{equation}
\hat{A}_j = \hat\Psi_{j}^\dagger(\text{\textbf{\textit{r}}}) \hat\Psi_{j}(\text{\textbf{\textit{r}}}),
\label{eq:dephasingop}
\end{equation}
with the associated system-bath interaction written as
\begin{equation}
\mathcal{\hat{L}}_j\left[\hat\rho\right] = 2\hat{A}_j\hat\rho\hat{A}_j^\dagger - \hat{A}_j^\dagger\hat{A}_j\hat\rho - \hat\rho\hat{A}_j^\dagger\hat{A}_j.
\label{eq:superopdephasing}
\end{equation}
Owing to the fact that the reservoirs are statistically independent, the loss term can be rewritten with an independent dephasing summand:
\begin{equation}
\begin{aligned}
	\!\!\!\!\left(\frac{\partial\hat\rho}{\partial t}\right)_{\!\text{loss}} \!\!\!&=\! \sum_{\text{\textbf{\textit{b}}}\in\mathbb{B}}{\mathcal{G}_{\text{\textbf{\textit{b}}}}\!\int \! \mathcal{\hat{L}}_{\text{\textbf{\textit{b}}}}\left[\hat\rho\right] \, d^3\text{\textbf{\textit{r}}}} + \!\sum^S_j{\mathcal{X}_j\!\int \! \mathcal{\hat{L}}_j\left[\hat\rho\right] \, d^3\text{\textbf{\textit{r}}}},\!\!\!\!\!
\end{aligned}
\label{eq:markovdephasing}
\end{equation}
where $S$ is the number of states. The last summation is over the set of reservoirs, each damping the phase of a dedicated state $j$ and characterized by a reservoir constant $\mathcal{X}_j$. Omitting the intermediate steps, the resulting equations for loss take the same algebraic form as Eqs.~(\ref{eq:mbatdensity}) except that the scalar $\kappa_1$ from Eq.~(\ref{eq:mbatdensity}) now absorbs the phenomenological dephasing rate $\gamma_d$:
\begin{equation}
	\kappa_1 = \frac{\gamma_{1}+\gamma_{2}+\gamma_d}{2}.
\label{eq:mbmasterlossconst2}
\end{equation}
The two reservoir constants have been absorbed into the pure dephasing rate by applying the rule $2(\mathcal{X}_1 + \mathcal{X}_2) \rightarrow \gamma_d$. The fact that $\gamma_d$ appears as a scaling factor to only $\langle \hat\Psi_i^\dagger\hat\Psi_j \rangle$ elements in Eqs.~(\ref{eq:mbloss}) with $i\neq j$ is physically justified: The state phase can only be destroyed by coupling to the degrees of freedom included in $R_{\text{ph}1} \oplus R_{\text{ph}2}$. This result suggests that the relative phase between states $1$ and $2$ can be manipulated by $\gamma_d$.

\section{Experimental decay rates}
\label{sec:popconversion}

Unlike in the equations above, many-body relaxation rates are usually not available in s$^{-1}$ units, because in an experiment one counts numbers of atoms rather than atomic densities given in density units, e.g., cm$^3$ s$^{-1}$ for two-body losses. Therefore a more practical version of Eqs.~(\ref{eq:mbatdensity}) should be found. The preferred form is where $\rho_{ii}$ are normalized populations. To convert Eqs.~(\ref{eq:mbatdensity}) to $\rho_{ij}$ the density equations are first integrated over the density profile, then conversion rules for decay constants are defined. The resulting equations for $N$ or $\rho$ lose explicit spatial dependence and describe the ensemble as a whole. They allow us to use the two-body decay rates in units of cm$^3$ s$^{-1}$. Under the assumptions of rapid rethermalization and constant temperature the cloud does not change its shape. The number rate equations read
\begin{equation}
  \begin{aligned}
	\frac{\partial N_1}{\partial t} &= - \gamma_1^{\text{th}} N_1 - k\gamma_{12}^{\text{th}} N_1 N_2 - k\gamma_{11}^{\text{th}} N_1^2, \\
	\frac{\partial N_2}{\partial t} &= - \gamma_2^{\text{th}} N_2 - k\gamma_{12}^{\text{th}} N_1 N_2 - k\gamma_{22}^{\text{th}} N_2^2, \\
	\frac{\partial N_{12}}{\partial t} &= - \kappa_1 N_{12} - \kappa_2 N_1 N_{12} - \kappa_3 N_{2} N_{12}, \\
	\frac{\partial N_{21}}{\partial t} &= - \kappa_1 N_{21} - \kappa_2 N_1 N_{21} - \kappa_3 N_{2} N_{21}
  \end{aligned}
\label{eq:numbermastereq}%
\end{equation}
with the auxiliary definitions where $\gamma_d$ is also included:
\begin{equation}
\begin{aligned}
	\!\!\!\!\kappa_1\!\! &=\! \frac{\gamma_{1}^{\text{th}}\!+\!\gamma_{2}^{\text{th}}\!+\!\gamma_d}{2},\,\,
	\kappa_2\!\! =\! k\frac{\gamma_{11}^{\text{th}}\!+\!\gamma_{12}^{\text{th}}}{2},\,\,
	\kappa_3\!\! =\! k\frac{\gamma_{22}^{\text{th}}\!+\!\gamma_{12}^{\text{th}}}{2}\!\!\!\!\!\!
\end{aligned}
\label{eq:mbmasterlossconstants}
\end{equation}
where $p \in \left\{x,y,z\right\}$, $k = (8 \pi^{3/2}\sigma_x\sigma_y\sigma_z)^{-1}$, and $\sigma_p=\sqrt{\frac{k_B T}{m \omega_p^2}}$ are the cloud widths, $\omega_p$ are the trap frequencies, $T$ is the ensemble temperature, $m$ is the atomic mass, and $k_B$ is the Boltzmann constant. The evolution of the ensemble density matrix is governed by the equation (later $\hbar = 1$)
\begin{equation}
	\frac{\partial\rho}{\partial t} = \frac{1}{i \hbar}\left[\mathbf{H}, \rho \right] + \left(\frac{\partial\rho}{\partial t}\right)_{\text{loss}},
\label{eq:mbmaster}
\end{equation}
where $\left(\left.\partial\rho\right/\partial t\right)_{\text{loss}}$ describes ensemble losses. After normalization of Eqs.~(\ref{eq:numbermastereq}) to $N_0$ the final ensemble density matrix elements with the unitary part of the ensemble master equation [Eq.~(\ref{eq:mbmaster})] constructed from the two-level Hamiltonian in the rotating wave approximation with one-body and two-body losses and pure dephasing become
\begin{equation}
  \begin{aligned}
	\!\!\!\!\!\frac{\partial \rho_{11}}{\partial t}\! &= \!- \Gamma_1 \rho_{11}\! -\! \Gamma_{12} \rho_{11} \rho_{22}\! -\! \Gamma_{11} \rho_{11}^2 \!+ \!\frac{i}{2} \Omega \left(\rho_{12}\!-\!\rho_{21}\right),\!  \\
	\!\!\!\!\!\frac{\partial \rho_{22}}{\partial t} \!&= \!- \Gamma_2 \rho_{22}\! -\! \Gamma_{12} \rho_{11} \rho_{22}\! - \!\Gamma_{22} \rho_{22}^2\! - \!\frac{i}{2} \Omega \left(\rho_{12}\!-\!\rho_{21}\right),\!  \\
	\!\!\!\!\!\frac{\partial \rho_{12}}{\partial t}\! &=\! - \varkappa_1 \rho_{12}\! - \varkappa_2 \rho_{11} \rho_{12} - \varkappa_3 \rho_{22} \rho_{12} \\
	&\quad+ \frac{i}{2} \Omega \left(\rho_{11}-\rho_{22}\right) + i \Delta\rho_{12},\!  \\
	\!\!\!\!\!\frac{\partial \rho_{21}}{\partial t}\! &=\! - \varkappa_1 \rho_{21}\! - \varkappa_2 \rho_{11} \rho_{21} - \varkappa_3 \rho_{22} \rho_{21} \\
	&\quad- \frac{i}{2} \Omega \left(\rho_{11}-\rho_{22}\right) - i \Delta\rho_{21}
  \end{aligned}
\label{eq:finalwithlosses}
\end{equation}
with the constants, where $\Gamma_d$ is the counterpart of $\gamma_d$:
\begin{equation}
	\!\!\!\varkappa_1\! =\! \frac{\Gamma_{1}+\Gamma_{2}+\Gamma_d}{2},\,
	\varkappa_2\! =\! \frac{\Gamma_{11}+\Gamma_{12}}{2},\,
	\varkappa_3\! =\! \frac{\Gamma_{22}+\Gamma_{12}}{2}.\!\!
\label{eq:lossconsts}
\end{equation}
$\Gamma_d$ plays the role of an extra degree of freedom to include relative phase offsets between the states. $\Gamma_d$ is not associated with the population loss and can incorporate miscellaneous experimental imperfections. The decay constants are obtained from the conversion, where $N_0 \equiv N(t=0)$:
\begin{equation}
  \begin{aligned}
\Gamma_1 &= \gamma_1^{\text{th}}, \quad\Gamma_{11} = k N_0 \gamma_{11}^{\text{th}},\\
\Gamma_{12} &= k N_0 \gamma_{12}^{\text{th}}, \quad\Gamma_{22} = k N_0 \gamma_{22}^{\text{th}},
  \end{aligned}
\label{eq:conversionrates}
\end{equation}
and to account for the multiplicity of the colliding particles \cite{Burt1997,Kagan} we define $\gamma_{\mathbf{\ell}}^{\text{th}}=M!\,\gamma_{\mathbf{\ell}}^{\text{BEC}}$, where $\mathbf{\ell}$ is a subscript for which $M!$ is computed as follows\footnoteremember{note1}{The $M!$ rule is experimentally confirmed up to $M=3$.}: $\mathbf{\ell} = 1$ or $\mathbf{\ell} = 2$ produce $M! = 1! = 1$, $\mathbf{\ell} = 11$ or $\mathbf{\ell} = 12$ or $\mathbf{\ell} = 22$ produce $M! = 2! = 2$, $\mathbf{\ell} = 111$ produces $M! = 3! = 6$. Note, that the one-body coefficients are the same in either representation, i.e., $\Gamma_1 = \gamma_1^{\text{th}} = \gamma_1^{\text{BEC}}$ and $\Gamma_2 = \gamma_2^{\text{th}} = \gamma_2^{\text{BEC}}$. Decay rates measured with condensed atoms are $M!$ times smaller than those measured with thermal atoms, where $M$ is the number of participating particles. This is attributed to the fact that the condensed particle wave functions overlap at the collision point resulting in unitary probability. Thermal-atom wave functions, by contrast, are not the same and at the collision point the probability turns out to be a factorial of the number of colliding particles due to bosonic bunching \cite{Burt1997,Kagan}.

\section{Discussion}

The Markov approximation is valid as long as the system exhibits short memory. To apply it in the present derivation we assumed that atoms do not return to the system once they experience an inelastic collision. Also, the bath correlation time for the damping process should be much shorter than the characteristic time scales of interest in the system, e.g., the inverse of coupling constants or the inverse of decay rates. Otherwise the bath correlations may be preserved and the Markov approximation breaks down.

To quantitatively assess the effect of three-body recombination, a suitable physical system can readily be implemented in spin-$1$ systems such as one of the stretched states of the hyperfine ground state of $^{87}$Rb, $F=1$. In this system two-body collisions are prohibited and for a thermal cloud with rapid rethermalization we arrive at the rate equation for the number of atoms $N(t)$, with $k_{111}=\left(2\pi \sqrt{3}\right)^{-3}\! \left(\sigma_x\sigma_y\sigma_z\right)^{-2}$:
\begin{equation}
	\frac{\partial N}{\partial t} = - \gamma_1 N - \gamma_{111} \, k_{111} \, N^3.
\label{eq:numbereq13}
\end{equation}
Integrating Eq.~(\ref{eq:numbereq13}) with respect to $t$ and choosing the physically justified solution gives the following exact result:
\begin{equation}
	\!\!N(t) = \left[\left(N_0^{-2}+k_{111}\frac{\gamma_{111}}{\gamma_1}\right)e^{2\gamma_1 t}-k_{111}\frac{\gamma_{111}}{\gamma_1}\right]^{-1/2}\!\!,
\label{eq:numbereq13sol}
\end{equation}
where $N_0\equiv N(0)$ is the initial population at $t=0$. It is particularly interesting that in $^{133}$Cs, e.g., it is possible to find a pair of levels for two-state interferometry that both experience intrastate collisions, that is, the corresponding two-body decay processes should not differ as much as in a system of one stretched state and one unstretched state. In such a system the visibility is enhanced. However, $\gamma_{111}$ is very small; the reported values measured in a BEC are in the order of $5.4\times 10^{-30}$~cm$^6$ s$^{-1}$ \cite{Mertes2007,Burt1997}. It follows from Eq.~(\ref{eq:numbereq13sol}) and lifetime measurements, e.g., in \cite{Ivannikov2013thesis}, that for typical experimental parameters the molecular formation due to three-body recombination is a negligible loss process which only becomes apparent at temperatures close to $T_{\text{crit}}$ for $N_0 > 10^6$, pressures an order of magnitude lower than the usual achievable (i.e., in the order of $10^{-12}$~Torr), and trapping times over minutes. The $N_0$ uncertainty expressed as the standard deviation of the atom number fluctuations $\sigma_N$ scales with $N_0$ and causes an increase of $\left.\Delta f\right/\!f$ via the collisional shift as $\Delta f_c = \left. 2\hbar\left(a_{22} - a_{11}\right)\sigma_N n_0\right/m$ practically limiting $N_0$ in trapped-atom clocks to units of $10^{4}$ \cite{Harber2002,Treutlein2008thesis}. Therefore, in most high-precision experiments with thermal atoms $\gamma_{111}$ can safely be disregarded.

Many-body collisions lead to the products of the ensemble density matrix elements in Eqs.~(\ref{eq:finalwithlosses}) that generally give nonexponential decays of populations \cite{Prentiss1988}. The model accounts for the collisions of atoms in different orientations through the decay constants for practicality: In this case extra constants arising from the angular momentum conservation algebra are absorbed in the decay rates. By examining the structure of the presented equations and their extended versions \cite{Ivannikov2013thesis} one can develop an \textit{informal} procedure to construct the loss part of the master equation under the same approximations for arbitrarily many participating bodies, provided the bunching coefficients are determined\footnoterecall{note1}, and bypass the direct invocation of quantum field theory. Using the relations of Eqs.~(\ref{eq:conversionrates}) the rates reported for thermal atoms and a condensate in s$^{-1}$ can readily be inserted to the model. The ensemble model also allows one to include phase noise by statistical averaging of $P_z$ over a given phase distribution what is noticeably more difficult to do via field theory than via Eqs.~(\ref{eq:finalwithlosses}).

It is important to incorporate one- and two-body losses into the master equation as the trapped-atom clock operates best in the ISRE regime: ISRE dramatically increases the coherence time making it possible to observe the influence of two-body losses on the Ramsey contrast at long evolution times \cite{Deutsch2010a}. The till-now reported ISRE model \cite{Deutsch2010a,Maineult2012,KleineBuning2011} has been used with two correction factors to two of the three model parameters: $1.6$ to $\Delta_0$ and $0.6$ to $\omega_{\text{ex}}$; the lateral elastic collisional rate $\gamma_c$ has not been corrected. The high data and data-fitting accuracies \cite{Deutsch2010a,KleineBuning2011} with yet very large correction factors are suggestive of an additional decoherence mechanism missing in the ISRE model. A Ramsey model governed by Eqs.~(\ref{eq:finalwithlosses}) may be able to bridge the gap by including the relevant collisional losses in the system. The two models provide a more comprehensive understanding of ensemble coherent dynamics over long evolution times in a trapped-atom clock, in particular regarding the interferometric contrast decay. The developed model should find its applications in Ramsey-type interferometries with trapped thermal ensembles. It can be verified in contemporary experimental systems exploring quantum dynamics.

\begin{acknowledgments}
The author thanks Bogdan Opanchuk for the explanations of his model \cite{Opanchuk2012a} and for helpful physics discussions.
\end{acknowledgments}


%

\end{document}